\documentclass[11pt]{article}
\usepackage{amsfonts}
\usepackage{amsmath,amssymb}
\usepackage{amscd}

\usepackage{indentfirst}
\newcommand{\hko}{\hookrightarrow}

\newcommand{\pa}{\partial}
\newcommand{\mnn}{\mathfrak{n}}
\newcommand{\n}{\nonumber\\}

\newcommand{\clp}{\cl_{p,q}}
\newcommand{\mm}{(-1)^{|\psi|}}
\newcommand{\ptt}{\phi_{(2)}}
\newcommand{\ptp}{\phi_{(3)}}
\newcommand{\ml}{\lie}
\newcommand{\lie}{\text{{\it\char'44}}}

\newcommand{\bec}{\begin{center}}
\newcommand{\eec}{\end{center}}

\newcommand{\bea}{\begin{array}}
\newcommand{\ear}{\end{array}}

\newcommand{\bfr}{\begin{flushright}}

\newcommand{\efr}{\end{flushright}}
\newcommand{\noi}{\noindent}\newcommand{\Ra}{\rightarrow}
\newcommand{\ri}{\rightarrow}
\newcommand{\me}{\frac{1}{2}}

\newcommand{\cl}{{\mt{C}}\ell}

\newcommand{\RR}{\mathbb{R}}\newcommand{\op}{\oplus}
\newcommand{\HH}{\mathbb{H}}
\newcommand{\ap}{\alpha}
\newcommand{\be}{\beta}
\newcommand{\ot}{\otimes}
\newcommand{\la}{\Lambda}
\newcommand{\bege}{\begin{equation}}
\newcommand{\enge}{\end{equation}}
\newcommand{\w}{\wedge}
\newcommand{\g}{\gamma}

\newcommand{\si}{\sigma}

\newcommand{\beq}{\begin{eqnarray}}\newcommand{\benu}{\begin{enumerate}}\newcommand{\enu}{\end{enumerate}}
\newcommand{\eeq}{\end{eqnarray}}
\newcommand{\mt}{\mathcal}

\newcommand{\vv}{{\bf v}}
\newcommand{\ee}{{\bf e}}
\newcommand{\uu}{{\bf u}}

\newcommand{\ww}{{\bf w}}
\newcommand{\CC}{\mathbb{C}}

\newcommand{\M}{\mt{M}}

\newcommand{\ZZ}{\mathbb{Z}}

\newcommand{\non}{\nonumber\\}

\newcommand{\OO}{\mathbb{O}}

\newcommand{\mk}{\mathfrak}

\newcommand{\clt}{{\mt{C}}\ell_{3,0}}
\newcommand{\cle}{{\mt{C}}\ell_{1,3}}

\newcommand{\bx}{\begin{pmatrix}}
\newcommand{\ex}{\end{pmatrix}}

\begin{document}
 \title{On Clifford Subalgebras, Spacetime Splittings and Applications}
 \author{{\bf R. da Rocha}\thanks{DRCC - Instituto de F\'{\i}sica Gleb Wataghin (IFGW),
 Unicamp, CP 6165, 13083-970, Campinas (SP), Brazil. 
E-mail: {\tt roldao@ifi.unicamp.br}.}
\and
{\bf J. Vaz, Jr.}\thanks{Departamento de Matem\'atica Aplicada, IMECC, Unicamp, CP 6065, 13083-859, Campinas (SP), Brazil. 
 E-mail: 
{\tt vaz@ime.unicamp.br}}}

\date{}\maketitle

\bigskip

\centerline{Abstract}
 \bigskip
{\footnotesize {$\ZZ_2$-gradings of  Clifford algebras are reviewed
and we shall be concerned with an $\ap$-grading based
on the structure of inner automorphisms, which is closely related to the spacetime splitting,
if we consider the standard conjugation map automorphism 
by an arbitrary, but fixed, splitting vector. 
After briefly sketching the orthogonal and parallel components of products of differential forms, where 
we introduce  the parallel [orthogonal] part as the space [time] component, 
we provide a detailed exposition  of the Dirac operator splitting and we show how the 
differential operator parallel and orthogonal components are related to the Lie derivative 
along the splitting vector and the angular momentum splitting bivector. We also introduce 
multivectorial-induced $\ap$-gradings and present the Dirac equation in terms of
the spacetime splitting, where the Dirac spinor field is shown to be a direct sum of
two quaternions. We point out some possible physical applications of the formalism developed. 
 }}\\
{}\bigskip\medbreak
\medbreak \noindent
\medbreak \noindent
Pacs numbers: 03.30.+p,  03.50.De, 03.65.Pm

\section{Introduction}

The class of $\ZZ_2$-gradings have a paramount importance in the investigations concerning, e.g., 
generalized $N$-extended supersymmetries,
since this formalism needs a graded Lie algebra to be constructed.
On the other hand, the relationship between spacetime splitting and Physics have been investigated in an wide context, e.g., Arnowitt-Deser-Misner (ADM)
 formalism \cite{arno}, and the splitting of the group SL(2,$\OO$) into the Lorentz group SL(2,$\CC$).  
The main purpose of this paper is to give a complete characterization concerning splitting in Clifford algebra, with the purpose of using $\ZZ_2$-gradings
in order to implement spacetime splittings in Clifford algebras, and to 
apply this formalism in order to obtain the most general form of Dirac equation from the inner automorphic $\ap$-grading viewpoint.

After presenting algebraic preliminaries on Clifford algebras in Sec. (\ref{1}), we revisit  in Sec. (\ref{two}) 
the formalism of $\ZZ_2$-gradings of Clifford algebras \cite{mdecs}. 
In Sec. (\ref{sert1}) spacetime splitting naturally arises from the inner automorphism framework via a particular 
chosen $\ap$-grading, where we show some properties of the orthogonal [temporal] and parallel [spatial] components 
of the Clifford product. After presenting the metric splitting in Sec. (5), in Sec. (\ref{ser5}) the splitting of the dual Hodge star operator is studied, and in Sec. (\ref{secdirac})
the Dirac operator is investigated from the point of view of the
parallel and orthogonal components of differential and codifferential operators. We show how these components are related to the Lie derivative 
along the 1-form field $n$ and to the angular momentum 2-form field. Besides, we completely investigate decompositions  
of the codifferential operator and also of the Hodge dual operator in terms of the proposed $\ap$-grading. 
In Secs. (\ref{cov}) and (\ref{seccod}) the covariant and Lie derivatives associated with the differential and codifferential
operators are respectively derived in the context of the inner automorphic $\ap$-grading. In Sec. (\ref{dde1})
we exhibit a dual equivalent inner automorphic $\ap$-grading. Relatively to the given 
$\ap$-grading  $\ap(\psi) = n\hat\psi n^{-1}$, responsible for the spacetime splitting in  
successive space-like manifold slices which are orthogonal to an observer worldline, there exists an equivalent dual decomposition  
in terms of the volume element associated with the $\ap$-even subalgebra.
The Dirac operator is calculated in each one of these two equivalent dual splittings.
In Sec. (\ref{autola}) multivectorial
inner automorphisms are introduced, generalizing the (1-)vetorial inner automorphisms presented, and we also exhibit all possible splittings of 
elements in $\cle$. Finally in 
Sec. (\ref{dididi}) Dirac equation splitting is presented in a straightforward manner. 
After briefly revisiting  Hestenes approach of Dirac theory \cite{hes67,hes75} --- where the usual even/odd splitting is used to
perform spacetime splittings (called \emph{projective splitting}) --- the Dirac equation  
associated with any $\ap$-grading \cite{mdecs} is investigated for 1-form fields $n=n(x)$, from the $\ap$-grading 
$\ap(\psi) = n\psi n^{-1}$ viewpoint. We show that the spinor field satisfying Dirac 
equation is described by the sum of two quaternions, besides writing Dirac equation in this context.  
We compare our result with the previous results in the literature, based on the $\ap$-gradings induced by standard, chiral (Weyl) and Majorana 
idempotents representations of $\CC\ot\cle$ 
in $\mathcal{M}(4,\CC)$ \cite{mdecs}.

\section{Preliminaries}
\label{1}
Let $V$ be a finite $n$-dimensional real vector space. We consider the tensor algebra $\bigoplus_{i=0}^\infty T^i(V)$ from which we
restrict our attention to the space $\Lambda(V) = \bigoplus_{k=0}^n\Lambda^k(V)$ of multivectors over $V$. $\Lambda^k(V)$
denotes the space of the antisymmetric
 $k$-tensors, isomorphic to the $k$-forms.  Given $\psi\in\Lambda(V)$, $\tilde\psi$ denotes the \emph{reversion}, 
 an algebra antiautomorphism
 given by $\tilde{\psi} = (-1)^{[k/2]}\psi$ ([$k$] denotes the integer part of $k$). $\hat\psi$ denotes 
the \emph{main automorphism or graded involution},  given by 
$\hat{\psi} = (-1)^k \psi$. The \emph{conjugation} is defined as the reversion followed by the main automorphism.
  If $V$ is endowed with a non-degenerate, symmetric, bilinear map $g: V\times V \rightarrow \RR$, it is 
possible to extend $g$ to $\la(V)$. Given $\psi=\uu_1\w\cdots\w \uu_k$ and $\phi=\vv_1\w\cdots\w \vv_l$, $\uu_i, \vv_j\in V$, one defines $g(\psi,\phi)
 = \det(g(\uu_i,\vv_j))$ if $k=l$ and $g(\psi,\phi)=0$ if $k\neq l$. Finally, the projection of a multivector $\psi= \psi_0 + \psi_1 + \cdots + \psi_n$,
 $\psi_k \in \la^k(V)$, on its $p$-vector part is given by $\langle\psi\rangle_p$ = $\psi_p$. 
The Clifford product between $\ww\in V$ and $\psi\in\la(V)$ is given by $\ww\psi = \ww\w \psi + \ww\lrcorner \psi$.
 The Grassmann algebra $(\la(V),g)$ 
endowed with the Clifford  product is denoted by $\cl(V,g)$ or $\cl_{p,q}$, the Clifford algebra associated with $V\simeq \RR^{p,q},\; p + q = n$.
In what follows $\RR,\CC$ and $\HH$ denote respectively the real, complex and quaternionic (scalar) fields.

\section{$\ap$-projections}
\label{two}
A vector space $V$ is said to be graded by an abelian group $G$  if it is expressible as a direct sum $V = \bigoplus_i V_i$ of subspaces 
labeled by elements $i\in G$. We say that the Clifford algebra $\cl(V,g)$ is graded by $G$ if its 
underlying vector space is a $G$-graded vector space and its product satisfies deg($ab$) = deg($a$) + deg($b$).
Heretofore we consider $G = \ZZ_2$.  The usual $\ZZ_2$-grading of $\cl(V,g)$ 
is given by $\cl(V,g) = \cl^+(V,g) \oplus \cl^-(V,g)$, where $\cl^{+(-)}(V,g)$ denotes the sum of all the even (odd)
 $k$-vectors. An arbitrary $\ZZ_2$-grading of $\cl(V,g)$ is given \cite{mdecs} by $\cl(V,g) = \cl_0\oplus\cl_1$, where the 
subspaces $\cl_i, i=0,1,$ satisfy $\cl_i\cl_j \subseteq \cl_{i+j(\rm{mod}\; 2)}$. To each decomposition there is an associated vector
space automorphism\footnote{It can be proved that $\ap$ is also an algebra isomorphism.}
 $\ap:\cl(V,g)\rightarrow\cl(V,g)$ defined by $\ap|_{\cl_i} = (-1)^i \rm{id}_{\cl_i}$. The projections $\pi_i$ on
$\cl_i$ are given by 
\bege\label{pro11}
\pi_i (\psi) = \me(\psi + (-1)^i \ap(\psi)).
\enge\noi In the sequel we denote $\pi_i(\psi)=\psi_i$. An element that belongs to 
 $\cl_0$ ($\cl_1$) is called $\ap$-$even$ ($\ap$-$odd$). With respect to the usual $\ZZ_2$-grading, $\cl_0 = \cl^+_{p,q}$,  
$\cl_1 = \cl^-_{p,q}$ and the grading automorphism is given by  $\widehat{(\;)}$.

\section{Spacetime splitting via inner automorphic $\ap$-gradings}
\label{sert1}

A vector space endowed with a constant signature ($p,q$) metric, isomorphic  to
 $\RR^{p,q}$, can be identified in a point $x\in M$ as the space  
$T_xM$ tangent to $M$ in this point, where $M$ is a manifold locally diffeomorphic to the (local) foliation   
 $\RR\times\Sigma$ of $M$. There always exist a 1-form time-like field $n$, normal to $\Sigma$, i.e., 
$n^2 = g(n,n) > 0$ and $g(n,\vec\vv) = 0,$ for all $\vec\vv\in T_x\Sigma$  \cite{baez}.
We denote the metric in $T_xM \simeq \RR^{p,q}$ by $g: \sec T_xM \times \sec T_xM\ri\RR$. The time-like 1-form field
 $n\in T_x^*M$ can be locally interpreted as being cotangent 
to the worldline of observer families, i.e., the dual reference frame relative velocity associated with such observers. 

The even subalgebra $\clp^+$  of $\cl_{p,q}$, associated with the graded involution  
is defined as $\cl_{p,q}^+ = \{\psi\in\cl_{p,q}\, |\, \ap(\psi):= \hat\psi = \psi\}$ and here we define the inner automorphic $\ap$-grading
 $\ap: \cl_{p,q}\rightarrow\cl_{p,q}$ as 
\bege\label{alfa}
{}{\ap(\psi) = n\hat\psi n^{-1}}\enge\noi 
From eq.(\ref{alfa}) it can be seen that  
the inner automorphism $\ap$ is invariant under dilations of $n$, and without loss of generality it is possible to 
consider 1-forms $n$ of unitary norm $n^2 = 1$. Eq.(\ref{alfa}) is then written as $\ap(\psi) = n\hat\psi n$.
Such $\ap$-grading is used until Sec. (\ref{seccod}).  

Given $\ee^0 \in T_x^*M$, it is well known that $n$ can be obtained via a Lorentz transformation $L$ from the formula
 $n(x) = L(x)\ee^0\tilde{L}(x)$, where $L$ is an element of the group\footnote{The group Spin$_+$($p,q$) is defined as being the group 
constituted by a product of an even number of unitary norm vectors in $\RR^{p,q}$.}  Spin$_+$(1,3), in the particular case of $\cle$.
In the more general case $L\in$ Spin$_+(p,q)$, a frame $\{\ee_p\}$ is said to be adapted to the observer if the following
conditions hold:
\benu
\item[i)] $n = L \ee_0 \tilde{L}$, 
\item[ii)] $n^\sharp(\ee_p) = 0$, 
\item[iii)] $\{\ee_p\}$ spans the local spacetime related to $n$. 
\enu\noi 

Note that the automorphisms 
$\hat\psi$ and $\ap(\psi)$  commute, which can be illustrated by the following diagram:

\[\begin{CD}
\clp @>\widehat{}>> \clp^\pm \\
@VV\ap V @VV\ap V\\
{\clp}_\|\,\,{\clp}_\bot @>\widehat{}>> {\clp}^\pm_\|\,\,,\,{\clp}^\pm_\bot
\end{CD}\]
\noi
Indeed,
\bege
\ap(\hat\psi) = n\psi n^{-1} = \widehat{n\hat\psi n^{-1}} = \widehat{\ap(\psi)}.
\enge\noi 
Among the vast possibilities to introduce  subspaces of $\clp$, we define 
\beq
{\clp^\| = \{\psi\in\clp\, |\, \ap(\psi) = n\hat\psi n = \psi\},\qquad
\clp^\bot = \{\psi\in\clp \,|\, \ap(\psi) = n\hat\psi n = -\psi\}}
\eeq\noi where $\clp^\|$ denotes the $\ap$-even (spatial) component of $\clp$, 
while $\clp^\bot$ denotes the $\ap$-odd  (temporal) component of $\clp$, which 
 is shown to be a $\ZZ_2$-graded algebra with respect to the inner automorphism given by eq.(\ref{alfa}). 
Indeed, given $\psi_\|$, $\phi_\|\in\cl_{p,q}^\|$ and 
$\psi_\bot$, $\phi_\bot\in\cl_{p,q}^\bot$, we have
\beq\label{roron}
n\widehat{\psi_\|\phi_\|} n &=&  {\psi_\|\phi_\|}\in \clp^\|,\nonumber\\
n\widehat{\psi_\|\phi_\bot} n &=&  -{\psi_\|\phi_\|}\in \clp^\bot,\nonumber\\
n\widehat{\psi_\bot\phi_\bot} n &=&  = {\psi_\|\phi_\|}\in \clp^\|
\eeq\noi The graded involution of  $\clp^\|$ is the same of  $\clp$.
The projectors $\pi_\|, \pi_\bot$, defined by relations
\bege\label{projc}
{}{\pi_\|(\psi)= \me(\psi + n\hat\psi n),\qquad\qquad \pi_\bot (\psi) = \me(\psi - n\hat\psi n)}\enge\noi
can be written as 
\beq\label{46}
\pi_\|(\psi) &=& n\cdot(n\w\psi) = (\psi\w n)\cdot n = \psi - n\w(n\cdot\psi),\nonumber\\
\pi_\bot(\psi) &=& n\w(n\cdot\psi) = (\psi\cdot n)\w n,\eeq\noi 
from which follows the relations
$
n\cdot \psi_\| =0,\qquad n\w\psi_\bot = 0.
$ 
Expressions in eq.(\ref{46}) have a full geometric motivation, since given $\uu, \vv\in \RR^{p,q}$ it follows that
\beq
\vv &=& \vv\uu\uu^{-1}= (\vv\cdot \uu + \vv\w \uu) \uu^{-1} = (\vv\cdot \uu) \uu^{-1} + (\vv\w \uu)\cdot \uu^{-1}\n &=& \vv_\| + \vv_\bot\eeq

The relations
\beq
&&\psi_\| = \pi_\|(\psi_\|),\quad \psi_\bot = \pi_\bot(\psi_\bot),\quad \pi_\|\pi_\bot = \pi_\bot\pi_\| = 0,\n
&&\pi_\|^2 = \pi_\|,\quad \pi_\bot^2 = \pi_\bot\quad {\rm and}\quad \pi_\bot + \pi_\| = 1
\eeq show explicitly that the operators in eq.(\ref{46}) are indeed projector operators. 
The convention used asserts that $\psi_\| \;[\psi_\bot]$ represents the spatial [temporal] component of $\psi\in\cl_{p,q}$. 

The identities
\bege
\psi_\| n = n\widehat{\psi_\|}, \qquad \psi_\bot n = -n{\widehat{\psi_\bot}}, \qquad n\psi n = {\widehat{\psi_\|}} - {\widehat{\psi_\bot}}
\enge\noi holds and shall be useful in what follows.

From eqs.(\ref{roron}) we see that $\clp^\|\clp^\| \hko \clp^\|, \;\;\clp^\|\clp^\bot \hko \clp^\bot$ and 
$\clp^\bot\clp^\bot \hko \clp^\|$. The Clifford product between 
two multivectors  $\psi,\phi\in\cl_{p,q}$ is given by 
\bege
\phi\psi = (\phi_\| + \phi_\bot)(\psi_\| + \psi_\bot) = \phi_\|\psi_\| + \phi_\bot\psi_\bot + \phi_\|\psi_\bot + \phi_\bot\psi_\|
\enge\noi where the parallel and orthogonal components of the Clifford product $\phi\psi$ are respectively given by:
\beq
(\phi\psi)_\| = \phi_\|\psi_\| + \phi_\bot\psi_\bot \in \clp^\|, \qquad 
(\phi\psi)_\bot = \phi_\|\psi_\bot + \phi_\bot\psi_\| \in \clp^\bot  
\eeq\noi
Then we have 
\beq
\quad(\vv\w\phi)_\| &=&\me(\vv\phi + \hat\phi\vv)_\|\n 
&=& \vv_\|\w\phi_\| + \vv_\bot\w\phi_\bot\eeq\noi But relations $\vv_\bot \w \phi_\bot = n\w(n\cdot\vv) \w n\w (n\cdot\phi_\bot) = 0$ follow 
from $n\w n = 0$. Therefore,
\bege
(\vv\w\phi)_\| = \vv_\|\w\phi_\|.
\enge\noi Analogously it is immediate to see that 
\bege
(\vv\w\phi)_\bot = \vv_\bot\w\phi_\| + \vv_\|\w\phi_\bot
\enge
\section{Metric projections}
\label{metric}
In order to express the metric $g^\|_x\equiv g^\|: T_x\Sigma \times T_x\Sigma \ri \RR$ in each point of the spatial manifold $\Sigma$, 
locally given by  $g^\|(\vv,\uu) = \vv_\|\cdot \uu_\|$, the spatial component  $g^\|$ of the spacetime  metric 
 $g:T_xM\times T_xM\ri\RR$ is obtained from the expression
 \beq
\vv_\|\uu_\| &=& \frac{1}{4}(\vv-n\vv n)(\uu - n\uu n)\n &=& \frac{1}{4}(\vv\uu - \vv n \uu n - n\vv n\uu + n\vv n^2\uu n).
\eeq \noi
 But using $\vv n = 2 \vv\cdot n  - n\vv$ (similarly for $\uu$),  
it follows that 
 \beq
g^\|(\uu,\vv)  &=& g(\uu,\vv) - (\vv\cdot n)(\uu\cdot n).\label{316}
\eeq
\noi If a local reference frame in $T_xM\simeq\RR^{p,q}$ is adopted, 
it is possible to express $n = n^\mu \ee_\mu, \vv = v^\mu \ee_\mu, \uu = u^\mu \ee_\mu$, and from eq.(\ref{316}) it follows that
\beq
g^\|_{\mu\nu} &=&  h_{\mu\nu} u^\mu v^\nu,\eeq\noi where $h_{\mu\nu} = g_{\mu\nu} - n_\mu n_\nu$.    
Then we have
\beq
g^\| &=& h_{\mu\nu}\ee^\mu\ot \ee^\nu = (g_{\mu\nu} - n_\mu n_\nu) \ee^\mu\ot \ee^\nu, \n
g^\bot &=&  n_\mu n_\nu \ee^\mu\ot \ee^\nu.\eeq\noi

\section{Decomposition of dual Hodge star operator}                                              
\label{ser5}
We denote pseudoscalars associated with  $\clp$ and $\clp^\|$ respectively by  
 $\eta\in\sec\la^{p+q}(T_x^*M)$ and $\tau\in\sec\la^{p+q-1}(T_x^*M)$. 
From the orientation of $\clp$ the orientation in $\clp^\|$ is defined as
\bege\label{orcl}
{}{\tau := n\cdot\eta}                                                            
  \enge \noi or equivalently, $\eta = n\w\tau.                                                                                                     
$ Eq.(\ref{orcl}) can  also be written as
\bege\label{orcl1}{}{\tau = n\eta}\enge \noi   and therefore relations   
\bege
n\tau = \hat{\tau} n\qquad{\rm and}\qquad \hat\eta = -n\hat\tau = -\tau n
\enge\noi immediately follow, since  $\tau\in\clp^\|$.

The dual Hodge star operator (DHSO) is defined in  $\cl_{p,q}^\|$ as
\beq\label{man}
\star_\|\vv_\| &=& \vv_\|\cdot\tau = \vv_\|\cdot(n\cdot\eta)\n
 &=& -n\cdot(\vv_\|\cdot\eta)
\eeq\noi From eqs.(\ref{46}), expressions
$\vv_\|\cdot\eta = \vv\cdot\eta - [n\w(n\cdot\vv)]\cdot\eta$ follows, and also 
\beq\label{pol1}
n\cdot(\vv_\|\cdot\eta) &=&n\cdot(\star\vv)\eeq\noi from which we conclude, 
using eqs.(\ref{man}) and (\ref{pol1}), that
$
\star_\|\vv_\| = -n\cdot(\star\vv).
$ We can also write
$
\star_\|\vv_\| = -n\cdot(\star\vv)_\bot
$ since $n\cdot (\star\vv)_\| = 0$. But $n\w(\star\vv)_\bot = 0$ and then the  relation 
$
\star_\|\vv_\| = -n(\star\vv)_\bot 
$ implies that
\bege
{}{(\star\vv)_\bot = -n(\star_\|\vv_\|)}
\enge\noi 
In the particular case of spacetime algebra $\cle$ where the $\ap$-grading is given by the graded involution, 
it is well known that $\cle^\| \simeq\clt$, 
and in the case of the $\ap$-grading given by eq.(\ref{alfa})  the $\ap$-even subalgebra of $\cle$ is given by 
$\cle^\|\simeq \cl_{0,3}\simeq\HH\op\HH$. More details are to be furnished in  Sec. (\ref{dididi}).
 
For multivectors $\psi\in\clp$ it follows that 
\beq\label{voy}
\star_\|\psi_\|&=&  n\cdot(\overline\psi_\|\cdot\eta)\eeq \noi Taking again the expression
 $\psi_\| = \psi - n\w(n\cdot\psi)$ we can express
\beq
\overline\psi_\| &=&  \star\hat\psi - [(\widetilde{n\cdot\tilde\psi})\w n]\cdot\eta
\eeq\noi Therefore $ n\cdot(\bar\psi\cdot\eta)= n\cdot(\star\hat\psi)$ and we can finally write from eq.(\ref{voy}) that 
${\star_\|\psi_\|=n\cdot(\star\hat\psi)}$ and using the properties $n\cdot(\;\;\;)_\| = 0$ and $n\w(\;\;\;)_\bot = 0$ we obtain
\beq
\star_\|\psi_\| &=& n(\star\hat\psi)_\bot\eeq\noi from where it follows that
$
(\star\psi)_\bot = n(\star_\|\hat\psi_\|)
$ and 
\beq
\star_\|\psi_\| &=&  n\cdot(\star\hat\psi)
\eeq\noi 
Also,
\beq \star_\|(n\cdot\psi) &=& (\widetilde{n\cdot \psi})\cdot \tau\n
&=& \widetilde{\psi_\bot}\cdot\eta
\eeq\noi and therefore
\bege
{}{\star_\|(n\cdot\psi) = \star(\psi_\bot)}
\enge

We now prove a similar expression for the parallel component $(\star\psi)_\|$ of $\star\psi$. 
According to relation $\widetilde{\psi_\bot} = \me(\tilde\psi - n\bar\psi n)$ we can  obviously write
\bege
\widetilde{\psi_\bot}\cdot\eta = \me(\tilde\psi - n\bar\psi n)\cdot\eta
\enge\noi Given $\psi_k\in\sec \Lambda^k(T_xM)$, it follows that  
\beq
(n\overline{\psi_k} n)\cdot\eta &=& \langle n\overline{\psi_k} n\eta\rangle_{n-k} = -  \langle\,n\overline{\psi_k} \hat\eta n\rangle_{n-k}\nonumber\\
 &=& -n(\widehat{\widetilde{\psi_k}\cdot \eta})n,
\eeq\noi and from expression
\beq
\widetilde{\psi_\bot}\cdot\eta &=& \me[(\tilde\psi\cdot\eta) + n(\widehat{\tilde\psi\cdot\eta}) n]\n
 &=& (\tilde\psi\cdot\eta)_\| = (\star\psi)_\|
\eeq
\noi the relation
\bege
{}{(\star\psi)_\| = \star(\psi_\bot) = \star_\|(n\cdot\psi)}
\enge\noi holds.

\section{Dirac operator splitting}
\label{secdirac}
Consider the Clifford fiber bundle over the manifold $M$,  denoted by  
 $\cl(M,g) := \bigcup_{x\in M} \cl(T_x M,g_x)$. Given  $n\in$ sec $T_x^*M$, the
local decomposition  $M = I\times \Sigma$ where $I$ is a interval of $\RR$, can be obtained if and only if $n\w dn = 0$, 
from Frobenius theorem. 
The Dirac operator is denoted by $\pa$,  and the differential $d \equiv \pa\w :$ $\sec\Lambda^k(T_x^*M)\ri\sec\Lambda^{k+1}(T_x^*M)$ 
and codifferential  $\delta \equiv -\pa\cdot:$  $\sec\Lambda^{k+1}(T_x^*M)\ri\sec\Lambda^{k}(T_x^*M)$ operators are defined in
$\cl(M,g)$ in terms of the Dirac operator. 
In what follows we find expression for the components $ d_\|, d_\bot, \delta_\|$ and 
$\delta_\bot$ and consequently, for $\pa_\|, \pa_\bot$. From relations  
$
(\psi\w\phi)_\| = \psi_\|\w\phi_\|
$ we have
\bege
(\psi\w\phi_\|)_\| = \psi_\|\w\phi_\|\quad{\rm and}\quad (\psi\w\phi_\bot)_\| = 0,
\enge\noi and from expression
\bege
(\psi\w\phi)_\bot = \psi_\|\w\phi_\bot + (\psi_\bot\w\phi_\|)_\|
\enge it is immediate that
\bege
(\psi\w\phi_\|)_\bot = \psi_\|\w\phi_\|,\qquad (\psi\w\phi_\bot)_\bot = \psi_\|\w\phi_\bot
\enge\noi We define the parallel $d_\|$ and orthogonal $d_\bot$ components of the differential operator $d$,
 from the following expressions
\bege
d_\|\psi_\| = (d\psi_\|)_\|, \quad d_\|\psi_\bot = (d\psi_\bot)_\bot,\quad d_\bot\psi_\| = (d\psi_\|)_\bot,\quad d_\bot\psi_\bot = 0,
\enge
\noi from where we obtain
\beq
d_\|\psi_\| &=& (d\psi_\|)_\|\n
 &=& n\cdot(n\w d\psi_\|),
\eeq
\beq
d_\bot\psi_\| &=& (d\psi_\|)_\bot = n\w(n\cdot d\psi_\|)\nonumber\\
&=& n\w\lie_n\psi_\|
\eeq\noi and
\beq
d_\bot\psi_\bot = (d\psi_\bot)_\| = n\cdot(n\w d\psi_\bot)\nonumber\\
&=& n\cdot\{n\w d[n\w(n\cdot\psi)]\}\nonumber\\
&=&  0 
\eeq\noi Also, the relation
\beq
d_\|\psi_\bot = (d\psi_\bot)_\bot &=& n\w(n\cdot d\psi_\bot)\nonumber\\
&=& n\w \{n\cdot[dn\w(n\cdot\psi) - n\w d(n\cdot\psi)]\}\nonumber\\
&=& - (n\cdot dn) \w \psi_\bot - n\w d_\|(n\cdot\psi)\nonumber\\
&=& n\w (\lie_n n) \w (n\cdot\psi) - n\w d_\|(n\cdot\psi)
\eeq is immediately derived. Now, define the multiform field $\Omega_\mu\in\sec\la(T_x^*M)$ as 
\bege
{}{\Omega_\mu = n\pa_\mu n = -(\pa_\mu n)n}\enge\noi 
We can show that $\Omega_\mu\in\sec \Lambda^2(T_x^*M)$, since $n\cdot \pa_\mu n = 0$ and $n^2 = 1$. It is well known
that the commutator computed between a  multivector and a bivector preserves the grading of the later, and we have 
\beq
\me[\Omega_\mu,\psi_k] &=& \langle\,n\pa_\mu n\psi_k\rangle_k\n
&=& -(\pa_\mu n)\cdot (n\w\psi_k) - (\pa_\mu n)\w (n\cdot\psi_k)
\eeq\noi Therefore
\beq\me [\Omega_\mu,\psi] &=&  -(\pa_\mu n)\cdot (n\w\psi) - (\pa_\mu n)\w (n\cdot\psi)\nonumber\\
&=& n\w(\pa_\mu n\cdot\psi) + n\cdot(\pa_\mu n\w\psi)
\eeq\noi where we used the relation $n\cdot\pa_\mu n = 0$. The commutator between $\Omega_\mu$ and $\psi_\|$ is given by 
\beq
\me [\Omega_\mu,\psi_\|] &=&  -(\pa_\mu n)\cdot (n\w\psi_\|) - (\pa_\mu n)\w (n\cdot\psi_\|)\nonumber\\
&=& n\w(\pa_\mu n\cdot\psi_\|)\nonumber\\
&=& - n\w(n\cdot\pa_\mu\psi_\|)
\eeq\noi where the identity $\pa_\mu n\cdot\psi_\| = -n\cdot\pa_\mu\psi_\|$ follows from relation $n\cdot\psi_\| = 0$. 
We also get the result
\bege
{}{\me[\Omega_\mu,\psi_\|] = -(\pa_\mu\psi_\|)_\bot}
\enge\noi since $n\w(n\cdot\;\;) = \pi_\bot(\;\;)$. On the other hand, we have
\beq
\me [\Omega_\mu,\psi_\bot] &=&  -(\pa_\mu n)\cdot (n\w\psi_\bot) - (\pa_\mu n)\w (n\cdot\psi_\bot)\nonumber\\
&=& - (\pa_\mu n)\w (n\cdot\psi_\bot)\nonumber\\
&=& -n\cdot(n\w\pa_\mu\psi_\bot),\eeq\noi and from the equivalence  $n\cdot(n\w\;\;) = \pi_\|(\;\;)$ it follows that
\bege
{}{\me [\Omega_\mu,\psi_\bot] = -(\pa_\mu\psi_\bot)_\|}\enge\noi   Also
\bege\label{dindi}
{}{\pa_\mu n = -\me[\Omega_\mu, n]}\enge\noi holds, since 
\bege
\me[\Omega_\mu, n] = \me[\Omega_\mu, n_\bot] = -(\pa_\mu n_\bot)_\| = -(\pa_\mu n)_\| = -\pa_\mu n.
\enge\noi We used the obvious property that $n = n_\bot$ and, from relation $n\cdot\pa_\mu n = 0$ it follows that 
\bege
(\pa_\mu n)_\bot = 0\qquad{\rm and}\qquad \pa_\mu n = (\pa_\mu n)_\|.
\enge Alternatively,  eq.(\ref{dindi}) follows from the 
 definition of $\Omega_\mu$. Indeed, \beq
\me[\Omega_\mu, n] &=& \me(n\pa_\mu nn - nn\pa_\mu n)\n
&=& - \pa_\mu n.
\eeq\noi 
From relations 
$
(\pa_\mu \psi_\|)_\bot  = -\me[\Omega_\mu, \psi_\|]$
 and
$
(\pa_\mu \psi_\bot)_\|  = -\me[\Omega_\mu, \psi_\bot]$
 we have
\beq
(\pa_\mu\psi_\|)_\| &=&\pa_\mu\psi_\| - (\pa_\mu\psi_\|)_\bot\nonumber\\
&=&\pa_\mu\psi_\| + \me [\Omega_\mu, \psi_\|]\eeq\noi and
\beq
(\pa_\mu\psi_\bot)_\bot &=&\pa_\mu\psi_\bot - (\pa_\mu\psi_\bot)_\|\nonumber\\
&=&\pa_\mu\psi_\bot + \me [\Omega_\mu, \psi_\bot]\eeq\noi

\section{Covariant and Lie derivative associated with the differential operator}
\label{cov}
Consider the orthogonal component of the differential operator acting on the parallel component of the multivector $\psi\in\cl_{p,q}$:
\beq
(d\psi_\|)_\bot &=& (\g^\mu\w\pa_\mu\psi_\|)_\bot\nonumber\\
&=& \g_\|^\mu\w(\pa_\mu\psi_\|)_\bot + \g_\bot^\mu\w(\pa_\mu\psi_\|)_\|\nonumber\\
&=& \g_\|^\mu\w (-\me[\Omega_\mu,\psi_\|])  + n\w n^\mu(\pa_\mu\psi_\| + \me[\Omega_\mu,\psi_\|]) \nonumber\\
&=& \g_\|^\mu\w n \w(n\cdot\pa_\mu\psi_\|) + n\w n^\mu\pa_\mu\psi_\| = n\w[\g^\mu(\pa_\mu n\cdot \psi_\|) + n^\mu\pa_\mu\psi_\|]\non
&=&n\w\lie_n \psi_\|
\eeq\noi where
\beq\label{selles}
\g_\bot^\mu &=& n\w(n\cdot \g^\mu) = nn^\mu,\n \g_\|^\mu &=& \g^\mu - n\w(n\cdot \g^\mu) = \g^\mu - nn^\mu\eeq\noi 
 and $\lie_n$ is the Lie derivative along $n$.  The expressions above are to be widely used in  Sec. (\ref{dididi}).

Defining the parallel component of the covariant derivative as
\bege
{}{D_\mu^\|\psi_\| = \pa_\mu\psi_\| + \me[\Omega_\mu,\psi_\|]}
\enge\noi we have
\beq
(d\psi_\|)_\| &=& (\g^\mu\w\pa_\mu\psi_\|)_\|\non
&=& \g_\|^\mu\w(\pa_\mu\psi_\|)_\| + \g_\bot^\mu(\pa_\mu\psi_\|)_\bot\non
&=& \g_\|^\mu\w(\pa_\mu\psi_\| + \me[\Omega_\mu,\psi_\|]) = \g_\|^\mu\w D_\mu^\|\psi_\|\non
&=&d_\|\psi_\|\nonumber
\eeq\noi
Therefore the relation  
$
{}{d_\| = \g_\|^\mu\w D_\mu^\|}
$ is immediately obtained. 

Now consider the  parallel component of the differential operator acting on the orthogonal component of the 
 multivector $\psi\in\cl_{p,q}$:
\beq
(d\psi_\bot)_\| &=& (\g^\mu\w\pa_\mu\psi_\bot)_\|\non
&=& \g_\|^\mu(-\me[\Omega_\mu,\psi_\bot])\label{ghj}\\
&=& n\cdot[n\w\g^\mu\w\pa_\mu n\w (n\cdot\psi_\bot)], \quad\text{since $n\cdot \pa_\mu n = 0$ and $n\cdot(n\cdot\psi_\bot) = 0$}\non
&=& n\cdot[n\w(\pa_\mu\w n)\w(n\cdot\psi_\bot)]\non
&=& 0.\label{ghk}\eeq\noi Finally, for the orthogonal component of $d\psi_\bot$ we have
\beq
(d\psi_\bot)_\bot &=&  (\g^\mu\w\pa_\mu\psi_\bot)_\bot\non
&=&\g_\|^\mu\w(\pa_\mu\psi_\bot)_\bot + \g_\bot^\mu(\pa_\mu\psi_\bot)_\|\non
&=&\g^\mu_\|\w(\pa_\mu\psi_\bot + \me[\Omega_\mu,\psi_\bot]) + \g_\bot^\mu\w(-\me[\Omega_\mu,\psi_\bot])\non
&=& \g_\|^\mu (D_\mu\psi_\bot) + n\w\lie_n n\w(n\cdot\psi_\bot)\nonumber
\eeq\noi But it is well known that $\g_\|^\mu\w D_\mu\psi_\bot = \g_\|^\mu\w(\pa_\mu\psi_\bot)
 + \g_\|^\mu \w\me[\Omega_\mu,\psi_\bot]$ and, from eqs.(\ref{ghj}) and
(\ref{ghk}) it follows that, $\g_\|^\mu(-\me[\Omega_\mu,\psi_\bot])= 0$. Therefore
\beq
\g_\|^\mu\w\pa_\mu\psi_\bot &=& \g_\|^\mu\w\{\pa_\mu[n\w(n\cdot\psi)]\}\non
&=& \g_\|^\mu \w [\pa_\mu n\w(n\cdot\psi) + n\w\pa_\mu(n\cdot\psi)]\non
&=& -n\w\g_\|^\mu \w\pa_\mu(n\cdot\psi)\eeq\noi But 
\beq
\me[\Omega_\mu,n\cdot\psi] &=& -n\w(\pa_\mu n \cdot (n\cdot\psi)) + n\cdot(\pa_\mu n\w(n\cdot\psi))\non
&=&-n\w(\pa_\mu n \cdot (n\cdot\psi)) -  (n\cdot \pa_\mu n)(n\cdot\psi) + \pa_\mu n\w[n\cdot(n\cdot\psi)]\non
&=&-n\w(\pa_\mu n \cdot (n\cdot\psi))  + \pa_\mu n\w[n\w(n\cdot\psi)]
\eeq\noi and then
$
n\w\me[\Omega_\mu,n\cdot\psi] = 0.
$
\noi Therefore it follows that
\beq
\g_\|^\mu\w\pa_\mu\psi_\bot &=& -n\w\g_\|^\mu \w\pa_\mu(n\cdot\psi)\non
&=& - n\w\g_\|^\mu\w D_\mu(n\cdot\psi)\eeq\noi Finally we have 
\bege\label{decoi}
{}{(d\psi_\bot)_\bot = -n\w\g_\|^\mu \w D_\mu (n\cdot\psi) + n\w\lie_n n\w(n\cdot\psi_\bot)}
\enge
\noi Using the definition of the operator $\pi_\bot$, we have 
\beq
\ml_n  [\pi_\bot(\psi)] &=&\ml_n n\w (n\cdot\psi) + n\w\ml_n(n\cdot\psi)
\eeq\noi On the other hand we know that
\beq 
\ml_n (n\cdot\psi) &=& n\cdot d(n\cdot\psi) + d[n\cdot(n\cdot\psi)]\non
&=& \ml_n n\w(n\cdot\psi) + n\w[n\cdot d(n\cdot\psi)]\non
&=& \ml_n n\w(n\cdot\psi) + n\w\{n\cdot[d(n\cdot\psi) + n\cdot(d\psi)]\},\; \text{since $n\cdot(n\cdot d\psi) = 0$}\non 
&=& \ml_n n\w(n\cdot\psi) + \pi_\bot(\ml_n\psi)\nonumber\eeq\noi
and therefore
\beq
\ml_n  [\pi_\bot(\psi)] &=&\ml_n n\w (n\cdot\psi) + n\w\ml_n(n\cdot\psi)\n
&=& \ml_n n\w (n\cdot\psi) + n\w\ml_n n\w(n\cdot\psi) + n\w\pi_\bot(\ml_n\psi)\eeq

\subsection{Necessary condition for $[\ml_n, d_\|] = 0$}

Since the Lie derivative $\psi\in\clp$ calculated along $n$
is given by
\beq
\ml_n d\psi &=& d(n\cdot d\psi) + n\cdot d(d\psi)\non
&=& d(n\cdot d\psi) + d[d(n\cdot\psi)]\non
&=& d\ml_n\psi
\eeq\noi and then  $d\ml_n = \ml_n d$, it follows that 
\beq
\ml_n d_\|\psi &=& \ml_n[n\cdot(n\w d\psi)]\non
&=& d(\ml_n\psi) - \ml_n n\w(n\cdot d\psi) - n\w \ml_n(n\cdot d\psi)\non
&=& d(\ml_n\psi) -  n\w\{d[n\cdot(n\cdot d\psi)]\} - n\w\{n\cdot d[d(n\cdot\psi)]\} - \ml_n n\w(n\cdot d\psi)\non
&=& d_\|(\ml_n\psi) - \ml_n n\w(n\cdot d\psi) \nonumber\eeq\noi Therefore
\bege
{}{\ml_n(d_\|\psi) = d_\|(\ml_n\psi) - \ml_n n\w(n\cdot d\psi)}
\enge showing that $\ml_n$ and $d_\|$ commute if and only if  $\ml_n n$ = 0, or equivalently when the acceleration
$\nabla_n n$ is identically null.

\section{Covariant and Lie  derivatives  related to codifferential operator}
\label{seccod}
Consider initially the parallel component $\delta_\|$ of the codifferential operator $\delta$. We define
\beq
\delta_\|\psi_\| &=& (\delta\psi_\|)_\|\n
 &=& -(\pa\cdot\psi_\|)_\|
\eeq\noi From the definition above we see that
\beq\label{comp}
(\pa\cdot\psi_\|)_\| &=& (\g^\mu\cdot \pa_\mu\psi_\|)_\|\non
&=& (\g^\mu_\|\cdot \pa_\mu\psi_\|)_\| + (\g^\mu_\bot\cdot \pa_\mu\psi_\|)_\bot\n
&=& \g^\mu_\|\cdot(D_\mu^\|\psi_\|) - n\cdot \left(\me[\Omega(n),\psi_\|]\right),\eeq\noi where
$\Omega(n):= n^\mu\Omega_\mu$ and the covariant derivative is given by 
 $D^\|_\mu\psi_\| = \pa_\mu\psi_\| + \me[\Omega_\mu,\psi_\|]$. On the other hand, 
\beq
\pa_\|\w(\star_\|\widehat{\psi}_\|)&=& \pa_\|\w[n(\star\widehat{\widehat{{\psi}}}_\bot)]\non
&=& d_\|[n(\star\psi)_\bot] = d_\|\w(n\widetilde\psi_\|\eta)\non
&=& \g^\mu_\|\w\left(\pa_\mu(n\tilde\psi_\|\eta) + \me[\Omega_\mu,n\tilde\psi_\|\eta]\right)\eeq\noi But from relations
\beq
[\Omega_\mu,n\widetilde\psi_\|\eta] &=& (\Omega_\mu n\widetilde\psi_\|\eta - n\widetilde\psi_\|\eta\Omega_\mu)\non
&=& [\Omega_\mu,n]\widetilde\psi_\|\eta + n[\Omega_\mu,\widetilde\psi_\|]\eta\eeq\noi we obtain, using 
 $\pa_\mu\eta = 0$, the expressions
\beq
\pa_\|\w(\star_\|\widehat{\psi_\|}) &=& \g_\|^\mu\w\left(\pa_\mu n)\widetilde{\psi_\|}\eta + n(\pa_\mu\widetilde{\psi_\|})\eta + \me[\Omega_\mu, n]\widetilde{\psi_\|}
\eta + \me
 n[\Omega_\mu,\widetilde{\psi_\|}]\eta\right]\non
&=& \g_\|^\mu\w[n(\pa_\mu\widetilde{\psi_\|})\eta + \me n[\Omega_\mu,\widetilde{\psi_\|}]\eta],\quad \text{since $\pa_\mu n = -\me[\Omega_\mu, n]$}\non
&=& \g^\mu_\|\w[n(D^\|_\mu\widetilde{\psi_\|})\eta]\non
&=& -\frac{n}{2}\left[\g^\mu_\|\pa_\mu\widetilde{\psi_\|} - (\widehat{\pa_\mu\widetilde{\psi_\|}})\g^\mu_\| + 
\left(\g_\|^\mu\me[\Omega_\mu,\widetilde{\psi_\|}]- \me\widehat{[\Omega_\mu,\widetilde{\psi_\|}]}
\g^\mu_\|\right)\eta\right]\non
&=& - n[\g^\mu_\|\cdot(D^\|_\mu\widetilde{\psi_\|})]\eta\eeq\noi Now the DHSO on the expression above is calculated:
\beq
\star_\|^{-1} [\pa_\|\w(\star_\|\widehat{\psi_\|})] &=& \tilde\tau^{-1} \{\widetilde{-[(\g^\mu_\|\cdot D^\|_\mu\widetilde{\psi_\|})n]\eta}\}\non
&=& -\tilde\tau^{-1}\tilde\eta[\widetilde{\g^\mu_\|\cdot(D^\|_\mu\widetilde{\psi_\|})}]n
= -n[(D_\mu^\|\psi_\|)\cdot\g^\mu_\|]n\non
&=& -\g_\|^\mu\cdot D_\mu^\|\psi_\|\eeq\noi Comparing this last expression with eq.(\ref{comp}), we obtain
\beq
(\pa\cdot\psi_\|)_\| 
&=& -\star^{-1}_\| d_\|\star_\|\widehat{\psi_\|} - n\cdot\left(\me[\Omega(n),\psi_\|]\right)\eeq\noi But as 
\beq
n\cdot\me[\Omega(n),\psi_\|] &=& n^\mu n\cdot\me[\Omega_\mu,\psi_\|]\non
&=& -n^\mu\{n\cdot [(\pa_\mu n)\cdot (n\w\psi_\|)]\}\non
&=& n^\mu\{(\pa_\mu n)\cdot [n\cdot (n\w\psi_\|)]\}\non
&=& \ml_n n\cdot\psi_\|\eeq\noi it follows that
\bege
{}{(\pa\cdot\psi_\|)_\| = -\star_\|^{-1}d_\|\star_\|\widehat{\psi_\|} - (\ml_n n)\cdot \psi_\|}
\enge

\section{Equivalent dual decompositions}
\label{dde1}We have defined from eq.(\ref{alfa}) the $\ap$-grading given by 
\bege\ap(\psi) = n\hat\psi n^{-1}\label{decalfa}
\enge\noi From eqs.(\ref{orcl}, \ref{orcl1}), as $\eta\in \sec\Lambda^{p+q}(T_x^*M)$ and therefore
 $\eta^2 = \pm 1$, it can be seen that
\beq 
\ap(\psi) &=& n\hat\psi n^{-1}\n
&=&  n\eta^2\hat\psi \eta^{-2}n^{-1}\n
&=& n\eta\psi\eta\eta^{-1}\eta^{-1}n^{-1}\n
&=& (-1)^{|\psi|(p+q-1)}\,\tau\psi\tau^{-1}\label{dectau}
\eeq\noi Therefore there is an equivalent decomposition of elements in $\clp$:
\benu
\item From the 1-form  field $n\in\sec T_x^*M$ in the normal bundle given by
\bege\label{alfag}{}{\ap(\psi)_n := n\hat\psi n^{-1}}
\enge\noi or 
\item Via the field defined by the volume element $\tau\in\sec \Lambda^{p+q-1}(T_x^*M)$ associated with
 $\clp^\|$, where the decomposition (given by eq.(\ref{alfag})) is now rewritten as
\bege {}{\ap(\psi)_\tau := (-1)^{|\psi|(p+q-1)}\,\tau\psi\tau^{-1}}
\enge\noi from  eq.(\ref{dectau}).
\enu

As decompositions given by  $\ap(\psi)_n$ and $\ap(\psi)_\tau$ are identical, it is immediate that the action of 
differential operators on such automorphisms are also equivalent. Indeed we explicitly calculate below the 
action of $d:\sec\la^p (T_x^*M) \Ra \sec\la^{p+1} (T_x^*M)$ in each one of the decompositions given by eqs.(\ref{decalfa})
and (\ref{dectau}):
\bege\label{111}
{}{d\ap(\psi)_n = dn\hat{\psi}n^{-1} - nd\hat\psi n^{-1} + (-1)^{|\psi|}n\hat\psi dn^{-1}}\enge\noi 

\bege\label{112}
{}{d\ap(\psi)_\tau = (-1)^{|\psi|(p+q-1)}\,(dn\hat{\psi}n^{-1} - nd\hat{\psi} n^{-1} - (-1)^{|\psi|}n\hat\psi dn^{-1})}\enge\noi 
Now, the action of the codifferential operator $\delta:\sec\la^p (T_x^*M) \Ra \sec\la^{p-1} (T_x^*M)$ in each one of the 
decompositions,  considering 
$\delta\phi = \star^{-1}d\star\hat\phi$ and $\star\phi = \tilde\phi\,\eta$ ($\phi\in\clp$), is given respectively by  
\beq
\delta\alpha(\psi)_n&=&\star^{-1}d\star\widehat{\ap(\psi)_n}\n
&=& \star^{-1}d\,(n\tilde\psi n^{-1}\eta)\n
&=& \star^{-1}\{dn^{-1}\,\tilde\psi n\eta - n^{-1}\,d\tilde\psi\,n\eta - \mm n^{-1}\tilde\psi\, dn \eta\}\n
&=& \be\{\eta n\psi(-dn^{-1})\eta - \eta n(-\widehat{d\psi})n^{-1}\eta - (-1)^{|\psi|}\eta(-dn)\psi n^{-1}\eta\}\n
&=& -(-1)^{(p+q-1)}\,\be\{n\hat\psi dn^{-1} + nd\psi n^{-1} + (-1)^{|\psi|}dn\hat\psi n^{-1}\},\label{113}
\eeq\noi where the constant  $\beta :=(-1)^{q + |\psi|(p+q-|\psi|)}$ arises from the property  $\star^{-1} = \beta\star$, and by 
\beq
\delta\alpha(\psi)_\tau&=&\star^{-1}d\star\widehat{\ap(\psi)_\tau}\n
&=& \star^{-1}d\,(n\eta\bar\psi \eta^{-1}n^{-1}\eta)\n
&=& \star^{-1}\{dn^{-1}\eta^{-1}\,\bar \psi n - n^{-1}\eta^{-1}\,d\bar\psi\,n - \mm n^{-1}\eta^{-1}\bar\psi\, dn \}\n
&=& \be\{ n\hat\psi\eta^{-1}(-dn^{-1})\eta -  n(-\widetilde{d\bar\psi})\eta^{-1}n^{-1}\eta - (-1)^{|\psi|}\eta(-dn)\bar\psi\eta^{-1} n^{-1}\eta\}\n
&=& (-1)^{|\psi|(p+q-1)}\,\be\{-n\hat\psi dn^{-1} + nd\psi n^{-1} - (-1)^{|\psi|}dn\hat\psi n^{-1}\}.\label{114}
\eeq\noi Since the Dirac operator $\pa$ is given by $d-\delta$, it follows from 
 eqs.(\ref{111}, \ref{112}, \ref{113}, \ref{114}), that 
$\pa(\ap(\psi)_n) = \pa(\ap(\psi)_\tau)$ can be written as
\bege\label{mussa}
{}{\pa(\ap(\psi)_n) = (1 + (-1)^{|\psi|}\be\Delta)dn\,\hat\psi\,n^{-1} - n(d\hat{\psi} - \be\Delta\,d\psi)
n^{-1} + (\be\Delta - (-1)^{|\psi|})n\hat\psi\,dn^{-1}}
\enge\noi where $\Delta:=(-1)^{|\psi|(p+q-1)}$. The expression above can be straightforwardly written in the particular case of Minkowski
spacetime as 
\bege\label{zaka}
{}{\pa(\ap(\psi)_n) = \pa(\ap(\psi)_\tau) = -2\left((-1)^{|\psi|}nd\psi n^{-1} + n\psi dn^{-1}\right)}
\enge \noi 
\section{$k$-vectorial inner automorphisms}
\label{autola}
Heretofore we treated inner automorphisms of $\clp$ given by 
\bege{\ap(\psi)_n := n\hat\psi n^{-1}}
\enge\noi where $n\in T_x^*M$ denotes a 1-form field. We can generalize these automorphisms, considering
more general automorphisms of $\clp\simeq \sec\cl(M,g)$ generated by  $k$-form fields instead of 
 1-form fields, from definition
\bege\label{ap11}
{}{\ap^{(k)}(\psi) := \phi_{(k)}\psi\phi_{(k)}^{-1}}
\enge\noi where $\phi_{(k)}\in\sec\Lambda^k(T_x^*M)$.
The projector operators given by eqs.(\ref{projc}) can be generalized by expressions
\bege\label{procje}
{}{\pi_\|^{(k)}(\psi)= \me(\psi + \phi_{(k)}\psi\phi_{(k)}^{-1} ),\qquad\qquad \pi_\bot^{(k)} (\psi) = \me(\psi - \phi_{(k)}\psi \phi_{(k)}^{-1})}\enge\noi

In the subsequent subsections we explicitly express  $\ap$-gradings for
all elements of $\cle$. Consider $\phi^{(k)}\neq\psi$, otherwise we have the trivial case $\ap^{(k)}(\psi) = \psi$.
In what follows consider  $1\leq i,j,k\leq 3$ distinct indices.
\subsection{Bivectorial inner automorphisms}
\label{autol}
In this case the decomposition is defined by 
\bege{}{\ap^{(2)}(\psi) := \ptt\psi \ptt^{-1}}
\enge\noi where $\ptt \in\sec\Lambda^2(T_x^*M)$ is a 2-form field.
We have the following cases ($i,j,k = 1,2,3$):
\benu
\item[a)] $\psi\in \la^1(T_x^*M)$:
\bec\begin{tabular}{||r||r||r|r|r||}\hline\hline
$\psi$& $\ptt$&$\ap^{(2)}(\psi)$&$\pi_\|^{(2)}(\psi)$&$\pi_\bot^{(2)}(\psi)$\\\hline\hline
$\ee^0$&$\ee^{0j}$&$-\ee^0$&0&$\ee^0$\\\hline
$\ee^0$&$\ee^{ij}$&$\ee^0$&$\ee^0$&0\\\hline
$\ee^i$&$\ee^{0j}$&$\ee^i$&$\ee^i$&0\\\hline
$\ee^i$&$\ee^{ik}$&$-\ee^i$&0&$\ee^i$\\\hline
$\ee^i$&$\ee^{0i}$&$-\ee^i$&0&$\ee^i$\\\hline
$\ee^i$&$\ee^{jk}$&$\ee^i$&$\ee^i$&0\\\hline
\hline\end{tabular}  \eec              
\noi
\item[b)] $\psi\in \la^2(T_x^*M)$:
\bec\begin{tabular}{||r||r||r|r|r||}\hline\hline
$\psi$& $\ptt$&$\ap^{(2)}(\psi)$&$\pi_\|^{(2)}(\psi)$&$\pi_\bot^{(2)}(\psi)$\\\hline\hline
$\ee^{0i}$&$\ee^{0j}$&$-\ee^{0i}$&0&$\ee^{0i}$\\\hline
$\ee^{0i}$&$\ee^{ij}$&$-\ee^{0i}$&0&$\ee^{0i}$\\\hline
$\ee^{0i}$&$\ee^{jk}$&$\ee^{0i}$&$\ee^{0i}$&0\\\hline
$\ee^{ij}$&$\ee^{0i}$&$-\ee^{ij}$&0&$\ee^{ij}$\\\hline
$\ee^{ij}$&$\ee^{0k}$&$\ee^{ij}$&$\ee^{ij}$&0\\\hline
$\ee^{ij}$&$\ee^{ik}$&$-\ee^{ij}$&0&$\ee^{ij}$\\\hline
\hline\end{tabular}  \eec              
\noi
\item[c)] $\psi\in \la^3(T_x^*M)$:
\bec\begin{tabular}{||r||r||r|r|r||}\hline\hline
$\psi$& $\ptt$&$\ap^{(2)}(\psi)$&$\pi_\|^{(2)}(\psi)$&$\pi_\bot^{(2)}(\psi)$\\\hline\hline
$\ee^{0ij}$&$\ee^{0j}$&$\ee^{0ij}$&$\ee^{0ij}$&0\\\hline
$\ee^{0ij}$&$\ee^{ij}$&$\ee^{0i}$&$\ee^{0ij}$&0\\\hline
$\ee^{0ij}$&$\ee^{jk}$&$-\ee^{0ij}$&0&$\ee^{0ij}$\\\hline
$\ee^{123}$&$\ee^{0i}$&$-\ee^{123}$&0&$\ee^{123}$\\\hline
$\ee^{123}$&$\ee^{ij}$&$\ee^{123}$&$\ee^{123}$&0\\\hline
\hline\end{tabular}                
\eec\noi
\item[d)] $\psi\in \la^4(T_x^*M)$:
\bec\begin{tabular}{||r||r||r|r|r||}\hline\hline
$\psi$& $\ptt$&$\ap^{(2)}(\psi)$&$\pi_\|^{(2)}(\psi)$&$\pi_\bot^{(2)}(\psi)$\\\hline\hline
$\ee^{0123}$&$\ee^{\mu\nu}$&$\ee^{0123}$&$\ee^{0123}$&0\\\hline
\hline\end{tabular}                
\eec\noi
\enu
\subsection{Trivectorial inner automorphisms}
Now the  $\ap$-grading is defined by 
\bege{}{\ap^{(3)}(\psi) := \ptp\psi \ptp^{-1}}
\enge\noi where $\ptp \in\sec\Lambda^3(T_x^*M)$ is a 3-form field.
\benu
\item[a)] $\psi\in \la^1(T_x^*M)$:
\bec\begin{tabular}{||r||r||r|r|r||}\hline\hline
$\psi$& $\ptp$&$\ap^{(3)}(\psi)$&$\pi_\|^{(3)}(\psi)$&$\pi_\bot^{(3)}(\psi)$\\\hline\hline
$\ee^0$&$\ee^{0ij}$&$\ee^0$&$\ee^0$&0\\\hline
$\ee^0$&$\ee^{123}$&$-\ee^0$&0&$\ee^0$\\\hline
$\ee^i$&$\ee^{0ij}$&$\ee^i$&$\ee^i$&0\\\hline
$\ee^i$&$\ee^{0jk}$&$-\ee^i$&0&$\ee^i$\\\hline
$\ee^i$&$\ee^{123}$&$\ee^i$&$\ee^i$&0\\\hline
\hline\end{tabular}  \eec              
\noi
\item[b)] $\psi\in \la^2(T_x^*M)$:
\bec\begin{tabular}{||r||r||r|r|r||}\hline\hline
$\psi$& $\ptp$&$\ap^{(3)}(\psi)$&$\pi_\|^{(3)}(\psi)$&$\pi_\bot^{(3)}(\psi)$\\\hline\hline
$\ee^{0i}$&$\ee^{0ij}$&$\ee^{0i}$&$\ee^{0i}$&0\\\hline
$\ee^{0i}$&$\ee^{123}$&$-\ee^{0i}$&0&$\ee^{0i}$\\\hline
$\ee^{ij}$&$\ee^{0ij}$&$\ee^{ij}$&$\ee^{ij}$&0\\\hline
$\ee^{ij}$&$\ee^{0ik}$&$-\ee^{ij}$&0&$\ee^{ij}$\\\hline
$\ee^{ij}$&$\ee^{123}$&$\ee^{ij}$&$\ee^{ij}$&0\\\hline
\hline\end{tabular}\eec              
\noi
\item[c)] $\psi\in \la^3(T_x^*M)$:
\bec\begin{tabular}{||r||r||r|r|r||}\hline\hline
$\psi$& $\ptp$&$\ap^{(3)}(\psi)$&$\pi_\|^{(3)}(\psi)$&$\pi_\bot^{(3)}(\psi)$\\\hline\hline
$\ee^{0ij}$&$\ee^{0ik}$&$\ee^{0ij}$&$\ee^{0ij}$&0\\\hline
$\ee^{0ij}$&$\ee^{123}$&$-\ee^{0ij}$&0&$\ee^{0ij}$\\\hline
$\ee^{123}$&$\ee^{0ij}$&$-\ee^{123}$&0&$-\ee^{123}$\\\hline
\hline\end{tabular}                
\eec\noi
\item[d)] $\psi\in \la^4(T_x^*M)$:
\bec\begin{tabular}{||r||r||r|r|r||}\hline\hline
$\psi$& $\ptp$&$\ap^{(3)}(\psi)$&$\pi_\|^{(3)}(\psi)$&$\pi_\bot^{(3)}(\psi)$\\\hline\hline
$\ee^{0123}$&$\ee^{\mu\nu\si}$&$-\ee^{0123}$&0&$\ee^{0123}$\\\hline
\hline\end{tabular}                
\eec\noi
\enu
\subsection{Tetravectorial inner automorphisms}
\label{autob} In this last case concerning $k$-vetorial decompositions in  $\cle$, 
the $\ap$-grading is defined by 
\bege{}{\ap^{(4)}(\psi) := \ee_{0123}\psi {\ee_{0123}}^{-1}}
\enge\noi where $\ee_{0123} \in\sec\Lambda^4(T_xM)$ denotes a volume element field in $\RR^{1,3}\simeq T_xM$.
All cases can be shown in the following table, for  $\psi\in \la^i(T_x^*M), i = 1,2,3$:
\bec\begin{tabular}{||r||r||r|r||}\hline\hline
$\psi$&$\ap^{(4)}(\psi)$&$\pi_\|^{(4)}(\psi)$&$\pi_\bot^{(4)}(\psi)$\\\hline\hline
$\ee^\mu$&$-\ee^\mu$&0&$\ee^\mu$\\\hline
$\ee^{\mu\nu}$&$\ee^{\mu\nu}$&$\ee^{\mu\nu}$&0\\\hline
$\ee^{\mu\nu\si}$&$-\ee^{\mu\nu\si}$&0&$\ee^{\mu\nu\si}$\\\hline
\hline\end{tabular}  \eec              
\noi This is clearly the case given by the grade involution $\ap^{(4)}(\psi) = \hat\psi$.
\subsection{Multivectorial inner automorphisms}
This case is simply the extension by linearity of the inner automorphism defined in Subsecs. (\ref{autol} - \ref{autob}), as
\bege
\ap^{\phi}(\psi):= \sum_{b=1}^4\,\ap^{(b)}(\psi)
\enge\noi where $\ap^{(1)}(\psi)\equiv \ap(\psi)_n = n\hat\psi n^{-1}$. We can also express
\bege{}{\ap^{\phi}(\psi) := \phi\psi\phi^{-1}},\qquad \phi, \psi\in\clp.\enge

\section{Dirac-Hestenes equation decomposition}
\label{dididi}
Hestenes approach to Dirac theory  \cite{hes67,hes75} is based on the real Clifford algebra $\cle$, 
instead of the Dirac algebra $\CC\ot\cle\simeq$ $\M$(4,$\CC$) used to formulate Dirac theory  
\cite{bjo,fel}. In such approach Hestenes asserts that a spinor field  $\psi$ is an element of the even subalgebra  $\cle^+$.

Dirac equation can be written as  \cite{bjo}
\bege\label{diracdef}
i\g_\mu\pa^\mu\psi = m\psi
\enge\noi where $\psi = (\psi_1, \psi_2, \psi_3, \psi_4)^\dagger$ is a column-vector in $\CC^4$. In order to represent 
vectors,
{spinors} and operators in a unique formalism, the  Clifford algebra, column-{spinors} are substituted by square matrices 
\cite{hes67,hes75,lou}. Using the standard representation of Dirac matrices, the idempotent 
\bege\label{idempo}
f = \me(1 + \ee_0)(1+i\ee_1\ee_2) \simeq{\begin{pmatrix}1&0&0&0\\0&0&0&0\\0&0&0&0\\0&0&0&0\end{pmatrix}}
\enge\noi is chosen in such a way that the {spinor} $\psi$ can be expressed as an algebraic {spinor}  \cite{fig,Fauser01}.
$$
\CC^4\ni\psi = {{\begin{pmatrix}\psi_1\\\psi_2\\\psi_3\\\psi_4\end{pmatrix} \simeq \begin{pmatrix}\psi_1&0&0&0\\\psi_2&0&0&0\\
\psi_3&0&0&0\\\psi_4&0&0&0\end{pmatrix}}} \in {\mt{M}}(4,\CC)f \simeq (\CC\ot\cle)f
$$\noi Since $i\psi = \psi\ee_1\ee_2$, the spinor $\Phi$ is defined in  $\cle\me(1+\ee_0)$. 

The automorphism $\alpha(\psi) = \ee_0 \hat{\psi} {\ee_0}^{-1}$, given by eq.(\ref{alfa}) 
in the particular case where $n = S\ee_0 S^{-1} = \ee_0$,  
induces a $\ZZ_2$-grading in $\cle$, given by  \cite{mdecs,Z2grad}:
\beq \cle^\|  &=& {\rm{span}}\;\{1, \ee_1, \ee_2, \ee_3, \ee_{12}, \ee_{23}, \ee_{31}, \ee_{123}\}\\
 \cle^\bot   &=& {\rm{span}}\;\{\ee_0,  \ee_{01}, \ee_{02}, \ee_{03}, \ee_{012},  \ee_{023},
                      \ee_{031}, \ee_{0123}\}
\eeq The subalgebra $\cle^+$ can be mapped in  $\cle^-$ if elements of $\cle^+ $ are multiplied by $\ee_0$. Indeed $\Phi$ 
can be written as \cite{lou}
$$
\Phi = \Phi_0 + \Phi_1 = (\Phi_0 + \Phi_1)\me(1 + \ee_0) = \me(\Phi_0 + \Phi_1\ee_0) + \me(\Phi_1 + \Phi_0\ee_0).
$$\noi We immediately verify that $\Phi_0 = \Phi_1\ee_0$ e $\Phi_1 = \Phi_0\ee_0$.
Taking the real part of eq.(\ref{diracdef}) 
we obtain
$$
\ee_\mu\pa^\mu \Phi\ee_2\ee_1 = m\Phi,\quad\Phi\in\cle\me(1 + \ee_0).
$$\noi The even component (associated with the graded involution) 
of the equation above with respect to the graded involution is the  Dirac-Hestenes equation \cite{hes75,lou}:
\bege\label{dh}
\pa\psi\ee_{1}\ee_{2} - m\psi \ee_0 = 0, \quad\psi\in\cl_{1,3}^+.
\enge\noi The spinor field $\psi: \RR^{1,3}\rightarrow \cle^+$  is denominated  Dirac-Hestenes spinor field (DHSF) 
 which is an operatorial {spinor}, since it generates the observables in Dirac theory. 
A DHSF  admits the canonical decomposition  \cite{lou}
$\psi = \sqrt\rho \exp(\ee_5\be/2)R$, denoting $\ee_5 = i\ee_0\ee_1\ee_2\ee_3$, if $\psi\tilde{\psi}\neq 0$, 
where $\sqrt\rho$ denotes a dilation, $\exp(\ee_5\be/2)$ is a dual rotation, $\be$ denotes the 
 Yvon-Takabayasi angle \cite{taka,yvon} and the bivector 
$R$, element of the group  Spin$_+$(1,3) $\simeq$ SL(2,$\CC$),
 is a Lorentz operator. 

Now, taking $n = S\ee^0 S^{-1}$ e $\si = S\ee^{12} S^{-1}$, where $S$ is a unitary operator, it is immediate that  
\bege\label{sicon}
n^2 = 1,\qquad \si^2 = -1, \qquad [n,\si]=0,\enge\noi  and from 
the idempotent given by eq.(\ref{idempo}) the idempotent $\me(1 + n)\me(1 + \si)$ can be constructed. 
On the other hand, under an arbitrary $\ap$-grading
it is well known that it is always possible to led Dirac-Hestenes equation (eq.(\ref{dh})) to the 
more general form given by  \cite{mdecs}
\bege\label{digen}
{}{\breve\pa\psi\si + m\psi n = 0}\qquad \psi\in\cl^\|
\enge\noi where $\breve\pa\psi = \pi_\|(\pa\psi) n + \pi_\bot(\pa\psi)$ and  $n$ is clearly $\ap$-odd, while $\si$ is $\ap$-even.

Considering the automorphism
 $\ap(\psi) = n\hat{\psi}n^{-1}$, it follows that given a spatial reference frame 
 $\{\mnn^1,\mnn^2,\mnn^3\}\in T^*_x\Sigma\simeq \RR^{0,3}$ adapted to $n := \mnn^0 = S\ee^0 S^{-1}$, 
such automorphism induces a  $\ZZ_2$-grading in $\cle$, given by  
\beq \cle^\|  &=& {\rm{span}}\;\{1, \mnn^1, \mnn^2, \mnn^3, \mnn^{12}, \mnn^{23}, \mnn^{31}, \mnn^{123}\},\label{pqp1}\\
 \cle^\bot &=& {\rm{span}}\;\{\mnn^0,  \mnn^{01}, \mnn^{02}, \mnn^{03}, \mnn^{012},  \mnn^{023},
                      \mnn^{031}, \mnn^{0123}\}.\label{pqp2}\eeq
\noi The 1-form fields $\mnn^i\in T_x^*\Sigma$ are exactly the fields  $\g_\|^i$ defined by eq.(\ref{selles}). 
As in the case of the $\ap$-grading given by $\ap(\psi) = n\psi n^{-1}$ the parallel component $\cle^\|$ of 
$\cle$ is given by  $\cle^\|\simeq \cl_{0,3}\simeq\HH\op\HH$, and then it is immediate to see that the spinor field composing 
eq.(\ref{digen}) is generated by
$\{{\tt P}_\pm, \mnn^{ij}{\tt P}_\pm\}$, where ${\tt P}_\pm = \me(1 \pm \mnn^{123})$ are idempotents spanned 
by central elements of $\cl_{0,3}$.
Each copy  $\HH$ of $\HH\op\HH\simeq\cl_{0,3}$ is respectively generated by  
$\{{\tt P}_+, \mnn^{ij}{\tt P}_+\}$ and $\{{\tt P}_-, \mnn^{ij}{\tt P}_-\}$, 
and therefore the spinorial field $\psi\in\cle^\|$ with respect to the
$\ap$-grading given by $\ap(\psi) = n\hat\psi n^{-1}$ has the more general form given by 
\bege\label{spin33}
{}{\psi = a{\tt P}_+ + b_{ij}\mnn^{ij}{\tt P}_+ + c{\tt P}_- + d_{ij}{\tt P}_-} \;\in \HH\op\HH\simeq\cl_{0,3} = \cle^\|,
\enge \noi where $a, b_{ij}, c, d_{ij}$ are scalar functions with values in $\CC$.

The possible values of $\si\in\cle^\|$ are given by the conditions in eq.(\ref{sicon}), and according to 
the elements  in $\cle^\|$ given by eq.(\ref{pqp1}) it follows immediately that  
\bege
{}{\si = a_3\mnn^{12} + a_1\mnn^{23} + a_2\mnn^{31}}\in {\rm SU}(2),
\enge\noi where $a_i\in\CC$, since from the condition $\si^2 = -1$ 
we have  $a_1^2 + a_2^2 + a_3^2 = 1$. Therefore  $\si\in$ SU(2) is a unitary quaternion. 
 
The parallel and orthogonal projection operators acting on the Dirac operator, in the case when the 
$\ap$-grading is given by $\ap(\psi)_n = n\hat\psi n^{-1}$ are defined by  \cite{mdecs}
\beq
\pi_\|(\pa) = \pi_\|(\mnn^\mu\pa_\mu):= \pi_\|(\mnn^\mu)\pa_\mu,\qquad
                 \pi_\bot(\pa) = \pi_\bot(\mnn^\mu\pa_\mu):= \pi_\bot(\mnn^\mu)\pa_\mu.
\eeq  
\noi We emphasize that $\mnn^0 \equiv n$.
Then eq.(\ref{digen}) can be written as
\beq
\mnn^k\pa_k\psi n\si + n\pa_n\psi\si + m\psi n = 0.
\eeq\noi Right multiplying the equation above by $n$ it follows that 
\bege
\mnn^k\pa_k\psi\si + n\pa_n\psi n\si + m\psi  = 0,
\enge\noi which is led to 
\bege\label{119}
{\mnn^k\pa_k\psi  + \pa_n\hat\psi =  m\psi \si}
\enge\noi This is the Dirac equation with respect to the $\ap$-grading given by $\ap(\psi) = n\hat\psi n^{-1}$, and motivated 
by spacetime decomposition. Spinor fields satisfying eq.(\ref{119}) are given by  eq.(\ref{spin33}). 

 \section{Concluding Remarks}
An arbitrary Clifford algebra (AC) is split in 
$\ap$-even and $\ap$-odd components, related to a given inner automorphic $\ap$-grading, 
besides describing various consequences of this decomposition in the splitting
of operators acting on the exterior and Clifford algebras. 
black hole. We use arbitrary AC $\ap$-gradings to generalize spacetime splitting  in a superposition 
of infinite spacelike slices, each one in a fixed time. 
Such a decomposition consists of a local spacetime foliation.  
Based on a $\ZZ_2$-grading, induced by a inner automorphism of $\clp$, we decompose  Dirac operator in parallel and orthogonal 
 components, showing how each of these components is related to the Lie derivative along 
the splitting vector. 
Parallel and  orthogonal projections are also introduced via inner automorphisms, 
 induced by multivectorial fields. In this form we introduce another $\ap$-grading completely dual and equivalent to the $\ap$-grading 
 $\ap(\psi) = n\hat\psi n^{-1}$, $\psi\in\clp$,
initially chosen. We have shown such dual $\ap$-grading is equivalent to the spacetime decomposition, but now
in terms of the volume element associated with the $\ap$-even subalgebra.
Dirac operator has been calculated in each of these dual decompositions.
Multivectorial-induced decompositions are calculated for all elements of $\cle$.
It can be shown that several symmetry breaking mechanisms in quantum field theory can be emulated using our formalism, 
such as  ${\mk{so}(7)\mapsto\mk{so}(6)\mapsto \mk{su}(3)\times\mk{u}(1)}$. 

The Dirac equation for an $\ap$-grading, associated with spacetime splitting via inner automorphisms, has been presented 
in a concise manner. Besides showing the Dirac spinor field is described by the direct sum of two quaternions, we have written
 Dirac equation, comparing our results with those recently obtained, also based on 
  $\ap$-gradings induced 
by the standard, chiral (Weyl) and Majorana representations
of  $\CC\ot\cle$ in $\mathcal{M}(4,\CC)$ \cite{mdecs}. 

We can show that relativistic kinematics and the evolution laws for the mass and spin of a test particle in the neighborhood
 of a Reissner-Nordstr\o m black hole are described via the formalism of AC splittings. 
The next step is to obtain these laws for a Kerr Black hole, extending all field equations --- based on Gauss-Codazzi equations 
 and in the relation between AdS$_5$ bulk geometry 
and the geometry of the 3-brane \cite{m} --- and formalizing these concepts to investigate physical effects of extra dimensions \cite{meu},
in a sequel paper.

\section{Acknowledgments}
The authors thank Prof. Waldyr  A. Rodrigues, Jr. and Dr. R. A. Mosna for useful suggestions.

\end{document}